\documentclass[11pt,a4paper]{article} 
\usepackage{amsthm}
\usepackage{amsmath}
\usepackage{amssymb}
\usepackage{bm}
\usepackage{verbatim}
\usepackage{color}
\usepackage{graphicx}
\usepackage[breaklinks]{hyperref} 

\definecolor{aocolour}{rgb}{0.7,0.8,1}
\definecolor{nncolour}{rgb}{1,0.8,0.7}

\newcommand{\set}[2]{\{\, {#1}\mid{#2}\,\}}
\newcommand{\setbig}[2]{\big\{ \: #1 \; \big| \; #2 \: \big\}}

\renewcommand{\emptyset}{\varnothing}
\renewcommand{\epsilon}{\varepsilon}

\newtheorem{definition}{Definition}

\newtheorem{theorem}{Theorem}
\newtheorem{lemma}{Lemma}

\newtheorem{claim}{Claim}
\numberwithin{claim}{lemma}

\newcommand{\wn}{\mathrm{wn}(\Sigma)}
\newcommand{\ewn}{\mathrm{wn_0}(\Sigma)}
\newcommand{\wnomega}{\mathrm{wn}^\omega(\Sigma)}
\newcommand{\wnID}{\mathrm{wnIDPDA}(\Sigma)}

\begin{document}

\sloppy

\title{Input-driven automata on well-nested infinite strings:
	automata-theoretic and topological properties\thanks{%
		This work was supported by the Russian Science Foundation, project 18-11-00100.}}
\author{Alexander Okhotin\thanks{%
	Department of Mathematics and Computer Science,
	St.~Petersburg State University, 7/9 Universitetskaya nab., Saint Petersburg 199034, Russia,
	\texttt{alexander.okhotin@spbu.ru}.}
	\and
	Victor L. Selivanov\thanks{%
	A.P. Ershov Institute of Informatics Systems, Novosibirsk, Russia,
	\emph{and}
	Department of Mathematics and Computer Science,
	St.~Petersburg State University,
	\texttt{vseliv@iis.nsk.su}.
	}}
\maketitle

\begin{abstract}
Automata operating on strings of nested brackets,
known as input-driven pushdown automata, and as visibly pushdown automata,
have been studied since the 1980s.
They were extended to the case of infinite strings
by Alur and Madhusudan (\href{http://dx.doi.org/10.1145/1007352.1007390}
	{``Visibly pushdown languages''}, STOC 2004).
This paper investigates the properties of these automata
under the assumption that a given infinite string is always well-nested.
This restriction enables a complete characterization
of the corresponding $\omega$-languages
in terms of classical $\omega$-regular languages
and input-driven automata on finite strings.
This characterization leads to a determinization result for these automata,
as well as to the first results on their Wadge degrees.  
\end{abstract}

\tableofcontents

\section{Introduction}

\emph{Input-driven pushdown automata} (IDPDA),
also known under the name of \emph{visibly pushdown automata},
are an important special class of pushdown automata,
introduced by Mehlhorn~\cite{Mehlhorn}.
In these automata, the input symbol determines
whether the automaton pushes a stack symbol,
pops a stack symbol
or does not access the stack at all.
These symbols are called
\emph{left brackets}, \emph{right brackets}
and \emph{neutral symbols},
and the symbol pushed at each left bracket
is always popped when reading the corresponding right bracket.
As shown by von Braunm\"uhl and Verbeek~\cite{vonBraunmuehl_Verbeek},
deterministic (DIDPDA)
and nondeterministic (NIDPDA) input-driven automata
are equivalent in power,
and the languages they recognize lie in the deterministic logarithmic space.
Input-driven automata enjoy excellent closure properties,
which almost rival those of finite automata,
and their complexity on terms of the number of states
has been extensively studied.
For more details, the reader is directed
to a survey by Okhotin and Salomaa~\cite{idpda_sigact_survey}.

A systematic study of input-driven automata was undertaken in a famous paper
by Alur and Madhusudan~\cite{AlurMadhusudan,AlurMadhusudan_JACM},
who, in particular,
generalized these automata to the case of infinite strings ($\omega$-IDPDA),
and identified them as a suitable model for verification.
By analogy with finite automata,
one can consider deterministic and nondeterministic $\omega$-IDPDA,
with B\"uchi and Muller acceptance conditions.
However, their properties are different:
for instance, unlike finite-state Muller $\omega$-automata,
Muller $\omega$-IDPDA cannot be determinized~\cite{AlurMadhusudan,AlurMadhusudan_JACM}.

This issue with determinization
was further studied by L\"oding et al.~\cite{LoedingMadhusudanSerre},
who discovered a more sophisticated acceptance condition,
under which determinization is possible.
The analysis made by L\"oding et al.~\cite{LoedingMadhusudanSerre} clearly indicates
that the main difficulty is caused by \emph{ill-nested inputs},
that is, infinite strings that may have \emph{unmatched left brackets}.
Recalling an important motivation for $\omega$-IDPDA
as a model of infinite computation traces
involving a stack~\cite{AlurMadhusudan,AlurMadhusudan_JACM},
an ill-nested input often means
a particular type of faulty behaviour: a memory leak.
A computation trace represented by a \emph{well-nested infinite string}
of the form $u_1 u_2 \ldots u_i \ldots$,
where each $u_i$ is a well-nested finite string,
is the normal behaviour.

This observation motivates the study
of a special class of $\omega$-IDPDA
operating on well-nested infinite strings.
These automata model a normal behaviour of an infinite process without memory leaks,
and they avoid the difficulties with determinization
discovered by L\"oding et al.~\cite{LoedingMadhusudanSerre}.

This paper investigates the properties of well-nested $\omega$-IDPDA,
which are defined in Section~\ref{idomega} in usual variants:
deterministic and nondeterministic,
with B\"uchi and Muller acceptance conditions.

In Section~\ref{section_characterization}, a characterization
of the corresponding $\omega$-IDPDA languages of well-nested strings
in terms of IDPDA on finite strings
and the usual $\omega$-regular languages is obtained.
This characterization leads to a determinization procedure for
well-nested Muller IDPDA,
as well as to the proof of their equivalence to nondeterministic B\"uchi IDPDA.

Besides the determinization question,
another property investigated by L\"oding et al.~\cite{LoedingMadhusudanSerre}
is the topology of $\omega$-languages
recognized by $\omega$-IDPDA.
These results motivate the study of Wadge degrees of $\omega$-IDPDA languages.
Although this collection is larger than that for the regular $\omega$-languages,
there is a hope to obtain an effective extension of Wagner hierarchy~\cite{wag}
to such languages which would nicely contrast
the case of (deterministic) context-free $\omega$-languages~\cite{fin,dup}.

The result of this paper, presented in Section~\ref{section_wadge},
is that the hierarchy of Wadge degrees of $\omega$-IDPDA languages of well-nested strings,
essentially coincides with the classical Wagner hierarchy of $\omega$-regular languages.

\section{Input-driven $\omega$-automata} \label{idomega}

Let $\Sigma$ be a finite alphabet.
The set of finite strings over $\Sigma$ is $\Sigma^*$,
the set of one-sided infinite strings is $\Sigma^\omega$.

In input-driven automata, the alphabet is
split into three disjoint sets
of \emph{left brackets} $\Sigma_{+1}$,
\emph{right brackets} $\Sigma_{-1}$
and \emph{neutral symbols} $\Sigma_0$.
In this paper, symbols from $\Sigma_{+1}$ and $\Sigma_{-1}$
shall be denoted by left and right angled brackets, respectively ($<$, $>$),
whereas lower-case Latin letters from the beginning of the alphabet ($a, b, c, \ldots$)
shall be used for symbols from $\Sigma_0$.

In a well-nested string,
any left bracket from $\Sigma_{+1}$ can match any right bracket from $\Sigma_{-1}$.
Formally, the set of well-nested finite strings, $\wn$,
is defined by the following grammar.
\begin{align*}
	S &\to SS
		\\
	S &\to {<} S {>}
		&& ({<} \in \Sigma_{+1}, \: {>} \in \Sigma_{-1})
		\\
	S &\to c
		&& (c \in \Sigma_0)
		\\
	S &\to \epsilon
\end{align*}
An \emph{elementary well-nested string}
is a well-nested string that is either a single symbol,
or a well-nested string enclosed in a pair of matching brackets:
$\ewn=\Sigma_0 \cup \Sigma_{+1} \wn \Sigma_{-1}$.

An infinite string $\alpha \in \Sigma^\omega$ is called \emph{well-nested},
if it is a concatenation of infinitely many elementary well-nested finite strings.
\begin{equation*}
	\wnomega = \ewn^\omega = \set{x_1 x_2 \ldots x_i \ldots}{x_i \in \ewn \text{ for all } i \geqslant 1}
\end{equation*}
Equivalently, an infinite string $\alpha$ is in $\wnomega$
if it has infinitely many well-nested prefixes.
Note that if the above definition had $x_i \in \wn$,
then it would also include all finite strings.

Input-driven automata on finite strings~\cite{Mehlhorn}
are pushdown automata with the following restriction on their use of the stack.
If the input symbol is a left bracket from $\Sigma_{+1}$,
then the automaton always pushes one symbol onto the stack.
For a right bracket from $\Sigma_{-1}$,
the automaton must pop one symbol.
Finally, for a neutral symbol in $\Sigma_0$,
the automaton may not access the stack at all.
The acceptance of a finite string
is determined by the state reached in the end of the computation,
whether it belongs to a set of accepting states $F \subseteq Q$.

The definition was extended to infinite strings
by adopting the notions of acceptance from finite $\omega$-automata.

\begin{definition}[Alur and Madhusudan~\cite{AlurMadhusudan}]\label{definition_idpda}
A nondeterministic input-driven pushdown automaton on infinite strings ($\omega$-NIDPDA)
consists of the following components.
\begin{itemize}
\item
	The input alphabet $\Sigma$
	is a finite set split into three disjoint classes:
	$\Sigma=\Sigma_{+1} \cup \Sigma_{-1} \cup \Sigma_0$.
\item
	The set of (internal) states $Q$
	is a finite set,
	with an initial state $q_0 \in Q$.
\item
	The stack alphabet $\Gamma$
	is a finite set,
	and a special symbol $\bot \notin \Gamma$
	is used to denote an empty stack.
\item
	For each neutral symbol $c \in \Sigma_0$,
	the transitions by this symbol
	are described by a function $\delta_c \colon Q \to 2^Q$.
\item
	For each left bracket symbol ${<} \in \Sigma_{+1}$,
	the behaviour of the automaton
	is described by a function $\delta_< \colon Q \to 2^{Q \times \Gamma}$,
	which, for a given current state, provides
	possible actions of the form
	``push a given stack symbol and enter a given state''.
\item
	For every right bracket symbol ${>} \in \Sigma_{-1}$,
	there is a function $\delta_> \colon Q \times \Gamma \to 2^Q$
	specifying possible next states,
	assuming that the given stack symbol is popped from the stack.
\item
	Acceptance is determined either by a B\"uchi condition,
	with a set of accepting states $F \subseteq Q$,
	or by a Muller condition,
	with a set of subsets $\mathcal{F} \subseteq 2^Q$.
\end{itemize}
A triple $(q, \alpha, x)$,
in which $q \in Q$ is the current state,
$\alpha \in \Sigma^\omega$ is the remaining input
and $x \in \Gamma^*$ is the stack contents,
is called a \emph{configuration}.
For each configuration, the next configuration is defined as follows.
\begin{align*}
	(q, c\alpha, x) &\vdash (r, \alpha, x)
		&& (q \in Q, \: c \in \Sigma_0, \: r \in \delta_c(q))
		\\
	(q, {<}\alpha, x) &\vdash (r, \alpha, sx),
		&& (q \in Q, \: {<} \in \Sigma_{+1}, \: (r, s) \in \delta_{<}(q))
		\\
	(q, {>}\alpha, sx) &\vdash (r, \alpha, x)
		&& (q \in Q, \: {>} \in \Sigma_{-1}, \: s \in \Gamma, \: r \in \delta_{>}(q, s))
\end{align*}

A \emph{run} on a well-nested string $\alpha \in \Sigma^\omega$
is any sequence of configurations $\rho = C_0, C_1, \ldots, C_i, \ldots$,
with $C_0=(q_0, \alpha, \epsilon)$
and $C_{i-1} \vdash C_i$ for all $i \geqslant 1$.
The set of states that occur in a run infinitely many times
is denoted by $\inf(\rho)$.

Under a B\"uchi acceptance condition,
an infinite string $\alpha$ is accepted,
if there exists a run
with at least one of the accepting states repeated infinitely often.
\begin{align*}
	L(\mathcal{B})
		&=
	\set{\alpha}{\exists \rho: \: \rho \text{ is a run on } \alpha, \: \inf(\rho) \cap F \neq \emptyset}
\intertext{%
Under a Muller acceptance condition,
the set of states repeated infinitely often
must be among the specified subsets.
}
	L(\mathcal{M})
		&=
	\set{\alpha}{\exists \rho: \: \rho \text{ is a run on } \alpha, \: \inf(\rho) \in \mathcal{F}}
\end{align*}
An automaton is deterministic,
if the cardinality of $\delta_c(q)$, $\delta_{<}(q)$ and $\delta_{>}(q, s)$
is at most 1 for any choice of arguments.
\end{definition}

Actually, the definition by Alur and Madhusudan~\cite{AlurMadhusudan}
further allows ill-nested infinite strings:
on unmatched right brackets,
it uses special transitions by an empty stack,
in which the automaton detects the stack emptiness
and leaves the stack empty;
on unmatched left brackets,
an automaton pushes symbols that shall never be popped.

In the case of finite strings,
the extension to ill-nested strings
does not affect the principal properties,
such as determinization and decidability.
For infinite strings, as it turns out, this detail makes a difference.
In the rest of this paper, all infinite strings are assumed to be well-nested.

\section{A characterization of $\omega$-input-driven languages of well-nested strings}\label{section_characterization}

The results of this paper are based on a characterization of $\omega$-input-driven languages
by $\omega$-regular languages,
which allows parts of the classical theory of $\omega$-regular languages
to be lifted to the case of well-nested infinite strings of brackets.

If an $\omega$-input-driven automaton $\mathcal{M}$ operates on a well-nested string,
this means that it reaches the bottom level of brackets infinitely often,
reading a well-nested string $x \in \ewn$ between every two consecutive visits.
The idea is to replace this string by a single symbol of a new alphabet,
which represents the essential information
on all possible computations of $\mathcal{M}$ on $x$.
Then the language of such encoded strings
shall be proved to be $\omega$-regular.

\begin{lemma}\label{muller_idpda_by_muller_regular_characterization_lemma}
Let $L \subseteq \wnomega$ be a language of well-nested infinite strings
over an alphabet $\Sigma=\Sigma_{+1} \cup \Sigma_{-1} \cup \Sigma_0$
recognized by an $n$-state deterministic input-driven Muller automaton.
Then there exists a finite set of pairwise disjoint languages,
$K_1, \ldots, K_m \subseteq \ewn$,
with $m \leqslant 2^{n^2} \cdot n^n$,
each recognized by an input-driven automaton,
and a regular $\omega$-language $M$ over the alphabet $\Omega=\{a_1, \ldots, a_m\}$
such that $L$ has the following representation.
\begin{equation*}
	L = \set{x_1 x_2 \ldots x_j \ldots}{x_j \in K_{i_j} \text{ for each } j \geqslant 0, \:
		a_{i_1} a_{i_2} \ldots a_{i_j} \ldots \in M}
\end{equation*}
The language $M$ is recognized
by a deterministic Muller automaton with $n \cdot 2^n$ states.
\end{lemma}
\begin{proof}
Let $\mathcal{M}=(\Sigma, Q, \Gamma, q_0, \langle\delta_\sigma\rangle_{\sigma \in \Sigma}, \mathcal{F})$
be a deterministic input-driven Muller automaton
that processes a string $x_1 x_2 \ldots x_\ell \ldots$.
The goal is to construct a deterministic finite Muller automaton $\mathcal{A}$
operating on a corresponding string $a^{(1)} a^{(2)} \ldots a^{(\ell)} \ldots$,
so that, after reading each prefix $a^{(1)} a^{(2)} \ldots a^{(\ell)}$,
the automaton $\mathcal{A}$ would know
the state reached by $\mathcal{M}$ after reading the prefix $x_1 x_2 \ldots x_\ell$,
as well as the set of all states passed through by $\mathcal{M}$
while reading the most recent well-nested substring $x_\ell$.
With this information, $\mathcal{A}$ can use its Muller acceptance condition
to simulate the Muller acceptance condition of $\mathcal{M}$.

The data to be computed for each well-nested substring $x \in \ewn$
is represented by a function $f_x \colon Q \to Q \times 2^Q$,
which maps a state $p$, in which $\mathcal{M}$ begins reading $x$,
to a pair $(q, S)$,
where $q$ is the state in which $\mathcal{M}$ finishes reading $q$,
and $S$ is the set of all states visited by $\mathcal{M}$ in this computation.
In the context of this proof, $f$ represents \emph{the behaviour of $\mathcal{M}$ on $x$.}

There are $m=2^{n^2} \cdot n^n$ different behaviour functions.
For each behaviour function $f$,
let $K_f \subseteq \ewn$ be the set of all strings,
on which $\mathcal{M}$ demonstrates the behaviour $f$.
A new Muller finite automaton promised in the statement of the lemma
shall be defined over an alphabet
$\Omega=\set{a_f}{f \colon Q \to Q \times 2^Q}$,
so that each symbol $a_f$
stands for any string $x \in K_f$.

Let $\rho$ be the run of $\mathcal{M}$ on $x_1 x_2 \ldots x_j \ldots$,
and denote by $q_{j,\ell}$ the state entered after reading $\ell$ first symbols of $x_j$.
\begin{equation*}
	\rho
		=
	\underbrace{q_{1,0}, q_{1,1}, \ldots, q_{1, |x_1|-1}}_{\text{reading } x_1},
	\underbrace{q_{2,0}, q_{2,1}, \ldots, q_{2, |x_2|-1}}_{\text{reading } x_2},
	q_{3, 1}, \ldots,
	\underbrace{q_{j,0}, q_{j,1}, \ldots, q_{q,|x_j|-1}}_{\text{reading } x_j},
	\ldots
\end{equation*}
For each substring $x_j$,
let $f_j=f_{x_j}$ be the behaviour function of $\mathcal{M}$ on this substring.
Then each $f_j$, applied to $q_{j,0}$,
provides the following data on the computation on $x_j$.
\begin{equation*}
	f_j(q_{j, 0})=\big(q_{j+1, 0}, \{q_{j,0}, q_{j,1}, \ldots, q_{j,|x_j|-1}, q_{j+1, 0}\}\big)
\end{equation*}

\begin{claim}\label{muller_idpda_by_muller_regular_characterization_lemma__acceptance_condition_claim}
A state $q$ occurs in the sequence $\{q_{j,\ell}\}$ infinitely many times
	if and only if
there exists a pair $(p, S)$, with $q \in S$,
which occurs as $\big(q_{j+1, 0}, \{q_{j,0}, q_{j,1}, \ldots, q_{j,|x_j|-1}, q_{j+1, 0}\}$
for infinitely many values of $j$.
\end{claim}
\begin{proof}
\begin{description}
\item[\textcircled{$\Rightarrow$}]
Assume that $q$ occurs in the run infinitely often.
Each time it occurs within some block $x_j$,
the corresponding pair
$(p, S)=\big(q_{j+1, 0}, \{q_{j,0}, q_{j,1}, \ldots, q_{j,|x_j|-1}, q_{j+1, 0}\}$
includes $q$ in the second component.
Since there are only finitely many pairs $(p, S)$,
at least one pair with the property $q \in S$ must occur infinitely many times.

\item[\textcircled{$\Leftarrow$}]
Let such a pair $(p, S)$, with $q \in S$,
occur as
$(p, S)=\big(q_{j+1, 0}, \{q_{j,0}, q_{j,1}, \ldots, q_{j,|x_j|-1}, q_{j+1, 0}\}$
for infinitely many $j$.
Every such occurrence indicates an occurrence of $q$ among the states $q_{j,\ell}$.
Then $q$ occurs in $\rho$ infinitely many times.
\end{description}
\end{proof}

Now the task is to simulate a run $\rho$ of $\mathcal{M}$ on $x_1 x_2 \ldots x_j \ldots$
by an automaton $\mathcal{M}'$
operating on the string $a_{f_1} a_{f_2} \ldots a_{f_j} \ldots$,
with every substring $x \in \ewn$ represented by a symbol $a_{f_x}$.

The desired Muller finite automaton $\mathcal{M}'$
is defined over the alphabet $\Omega=\set{a_f}{f \colon Q \to Q \times 2^Q}$
as follows.
Its states are pairs $(q, S)$,
where $q$ shall be the state in the corresponding run of $\mathcal{M}$ at this point,
whereas $S$ shall be the set of states visited by $\mathcal{M}$
on the last well-nested substring.
\begin{align*}
	P &= \set{(q, S)}{q \in Q, \: S \subseteq Q}
\intertext{%
A transition by $a_f$ applies the behaviour function $f$ to $q$
to determine the next state of $\mathcal{M}$,
as well as the set of states visited on the next well-nested substring.
}
	\Delta((q, S), a_f) &= f(q)
\end{align*}

\begin{claim}
In the run $\rho'$ of $\mathcal{M}'$ on $a_{f_1} a_{f_2} \ldots a_{f_j} \ldots$,
the state reached after reading each $a_{f_j}$
is $\big(q_{j+1, 0}, \{q_{j,0}, q_{j,1}, \ldots, q_{j,|x_j|-1}, q_{j+1, 0}\}\big)$.
\end{claim}

Putting together this property
and Claim~\ref{muller_idpda_by_muller_regular_characterization_lemma__acceptance_condition_claim},
a state $q$ occurs in $\rho$ infinitely often
	if and only if
there exists a pair $(p, S)$, with $q \in S$,
which occurs in $\rho'$ infinitely often.
\begin{equation*}
	\inf \rho = \bigcup_{(p, S) \in \inf \rho'} S
\end{equation*}

Then the Muller acceptance condition for $\mathcal{M'}$
is defined by putting together all states of $\mathcal{M}$
visited infinitely often and checking whether this set is in $\mathcal{F}$.
\begin{equation*}
	\mathcal{G} = \setbig{R}{R \subseteq Q \times 2^Q, \: \bigcup_{(q, S) \in R} S \in \mathcal{F}}
\end{equation*}

\begin{claim}
Let $x_1 x_2 \ldots x_j \ldots$, with $x_j \in \ewn$,
be an infinite string,
and let $f_j \colon Q \to Q \times 2^Q$ be the behaviour function on each $x_j$.
Then $\mathcal{M}$ accepts $x_1 x_2 \ldots x_j \ldots$
	if and only if
$\mathcal{M}'$ accepts $a_{f_1} a_{f_2} \ldots a_{f_j} \ldots$.
\end{claim}

In order to complete the proof of the lemma,
it remains to prove that
each language $K_f \subseteq \ewn$ of finite strings
is recognized by an input-driven automaton.

\begin{claim}
There exists a deterministic input-driven automaton $\mathcal{A}$
with the set of states
$Q'=\{q'_0\} \cup \set{(f, b)}{f \colon Q \to Q \times 2^Q, \: b \in \{0, 1\}}$,
which, upon reading a well-nested string $x \in \ewn$,
enters a state $(f, 0)$,
where $f \colon Q \to Q \times 2^Q$
is the behaviour function of $\mathcal{M}$ on $x$.
\end{claim}
\begin{proof}
Besides computing the behaviour,
the automaton also has to verify that the input string is in $\ewn$,
and for that reason it has to remember whether it is currently inside any brackets.
In a state $(f, b)$,
the first component is the desired behaviour function
for the longest well-nested suffix $v \in \ewn$,
whereas the second component indicates
whether the automaton is currently at the outer level of brackets ($0$) or not ($1$).
No transitions in a state of the form $(f, 0)$ are possible.
The stack alphabet is
$\Gamma'=\set{(f, <)}{f \colon Q \to Q \times 2^Q, \: {<} \in \Sigma_{+1}} \cup \Sigma_{+1}$,
where a symbol of the form $(f, <)$ is pushed while inside the brackets,
whereas on the outer level of brackets, the automaton pushes just a left bracket.

Let $f_\epsilon \colon Q \to Q \times 2^Q$
be the behaviour function on the empty string,
defined by $f(q)=(q, \{q\})$ for each $q \in Q$.
For each neutral symbol $c \in \Sigma_0$,
the behaviour on this symbol, $f_c \colon Q \to Q \times 2^Q$,
is defined by $f(q)=(\delta_c(q), \{q, \delta_c(q)\})$.
If $f$ is the behaviour function on $u \in \wn$,
and $g$ is the behaviour function on $v \in \wn$,
then the behaviour function on $uv$
is the following kind of function composition
that accummulates the states visited.
\begin{align*}
	(g \circ f)(p)=(r, \: S \cup T),
		&& \text{ where } f(p)=(q, S), \; g(q)=(r, T)
\end{align*}

The initial state of $\mathcal{A}$ is $\widehat{q}_0$;
upon reading a neutral symbol $c \in \Sigma_0$ in the initial state,
the automaton enters the state corresponding to the desired behaviour on $c$.
\begin{equation*}
	\delta'_c(q'_0)=(f_c, 0)
\end{equation*}
If the automaton reads a left bracket ${<} \in \Sigma_{+1}$ in the initial state,
it begins constructing a new behaviour function on the inner level,
remembering the bracket in the stack.
\begin{equation*}
	\delta'_<\big((f, 0)\big)=\big((f_\epsilon, 1), {<}\big)
\end{equation*}
When a matching right bracket ${>} \in \Sigma_{-1}$ is read,
the automaton pops the left bracket ${<} \in \Sigma_{+1}$ from the stack,
and therefore knows that it is returning to the outer level of brackets.
Then it merges the computed behaviour on the substring inside the brackets
with the actions made on the outer pair of brackets.
\begin{align*}
	\delta'_>\big((f, 1), {<}\big)=(f', 0),
		&& \text{ where } f'(p)=(r, S \cup \{p, r\}), \:
		\delta_<(p)=(q, s), \: r=\delta_>(f(q), s)
\end{align*}
Inside the brackets, upon reading a neutral symbol $c \in \Sigma_0$,
behaviour functions are composed as follows.
\begin{equation*}
	\delta'_c\big((f, 1)\big)=f_c \circ f
\end{equation*}
When a left bracket ${<} \in \Sigma_{+1}$ is encountered inside the brackets,
a new simulation is started inside,
but the current behaviour function is stored in the stack.
\begin{equation*}
	\delta'_<\big((f, 1)\big)=\big((f_\epsilon, 1), (f, {<})\big)
\end{equation*}
Upon reading a matching right bracket ${>} \in \Sigma_{-1}$,
the two simulations are composed.
\begin{align*}
	\delta'_>\big((g, 1), (f, {<})\big)=(h \circ f, \: 1),
		&& \text{ where } h(p)=(r, S \cup \{p, r\}), \:
		\delta_<(p)=(q, s), \: r=\delta_>(f(q), s)
\end{align*}
The transitions of the automaton $\mathcal{A}$ have been defined.
For each behaviour function $f \colon Q \to Q \times 2^Q$,
setting the set of accepting states to be $F_f=\{(f, 0)\}$
yields the desired IDPDA recognizing the language $K_f$
of all strings in $\ewn$, on which $\mathcal{M}$ behaves as specified by $f$.
\end{proof}

This completes the proof of the lemma.
\end{proof}

The same characterization as in Lemma~\ref{muller_idpda_by_muller_regular_characterization_lemma}
also extends to nondeterministic input-driven Muller automata,
at the expense in using exponentially more types of well-nested strings.

\begin{lemma}\label{muller_nidpda_by_muller_regular_characterization_lemma}
Let $L \subseteq \wnomega$ be a language of well-nested infinite strings
over an alphabet $\Sigma=\Sigma_{+1} \cup \Sigma_{-1} \cup \Sigma_0$
recognized by an $n$-state nondeterministic input-driven Muller automaton.
Then there exists a finite set of pairwise disjoint languages,
$K_1, \ldots, K_m \subseteq \ewn$,
with $m \leqslant 2^{n^2 \cdot 2^n}$,
each recognized by an input-driven automaton,
and a regular $\omega$-language $M$ over the alphabet $\Omega=\{a_1, \ldots, a_m\}$
such that $L$ has the following representation.
\begin{equation*}
	L = \set{x_1 x_2 \ldots x_j \ldots}{x_j \in K_{i_j} \text{ for each } j \geqslant 0, \:
		a_{i_1} a_{i_2} \ldots a_{i_j} \ldots \in M}
\end{equation*}
The language $M$ is recognized
by a nondeterministic Muller automaton with $n \cdot 2^n$ states.
\end{lemma}
\begin{proof}[Sketch of a proof.]
The construction differs from that in Lemma~\ref{muller_idpda_by_muller_regular_characterization_lemma}
in two respects.

First, since the original input-driven Muller automaton is now nondeterministic,
its behaviour on a well-nested substring $x \in \ewn$
can no longer be characterized
by an object as simple as a function $f \colon Q \to Q \times 2^Q$
describing the computation on $x$ beginning in each state.
For a nondeterministic automaton, a computation on $x$
may begin in any state $p$, pass through any set of states $S \subseteq Q$,
and finish reading $x$ in any state $q$:
this is a triple $(p, S, q)$.
The automaton's behaviour on $x$
is a \emph{set of possible triples},
or a \emph{ternary relation} $R \subseteq Q \times 2^Q \times Q$.
This accounts for the increase in the alphabet size $m$.

Second, besides the nondeterminism exercised while reading a substring $x$,
the nondeterministic input-driven Muller automaton
can also behave nondeterministically
when proceeding from one well-nested substring to another.
This nondeterminism is directly simulated
by the nondeterministic finite Muller automaton constructed in the lemma.
\end{proof}

The representation of well-nested $\omega$-input-driven languages
by $\omega$-regular languages
given in Lemmata~\ref{muller_idpda_by_muller_regular_characterization_lemma}--\ref{muller_nidpda_by_muller_regular_characterization_lemma}
has the following converse representation.

\begin{lemma}\label{muller_regular_characterization_to_muller_didpda_lemma}
Let $K_1, \ldots, K_m \subseteq \ewn$ be pairwise disjoint languages,
each recognized by an input-driven automaton,
and let $M$ be a regular $\omega$-language over the alphabet $\Omega=\{a_1, \ldots, a_m\}$.
Then, the following language is recognized by a deterministic input-driven Muller automaton.
\begin{equation*}
	L = \set{x_1 x_2 \ldots x_j \ldots}{x_j \in K_{i_j} \text{ for each } j \geqslant 0, \:
		a_{i_1} a_{i_2} \ldots a_{i_j} \ldots \in M}
\end{equation*}
\end{lemma}
\begin{proof}
Let each $K_i$ be recognized by a DIDPDA
$\mathcal{A}_i=(\Sigma, Q_i, \Gamma_i, q^{(i)}_0, \langle\delta^{(i)}\rangle_{c \in \Sigma}, F_i)$,
and let $\mathcal{M}=(\Omega, P, p_0, \eta, \mathcal{F})$
be a deterministic Muller automaton recognizing $M$.
A new deterministic input-driven Muller automaton $\mathcal{M}'$
uses the following set of states.
\begin{align*}
	Q'
		&=
	P
	\cup
	\underbrace{(Q_1 \times \ldots \times Q_m)}_{\text{used inside brackets}}
	\cup
	\{p_{dead}\}
	\\
	\Gamma'
		&=
	(P \times \Sigma_{+1})
	\cup
	(\Gamma_1 \times \ldots \times \Gamma_m)
\end{align*}
The states from $P$ are used outside all brackets.
The initial state is $p_0$.
Whenever the automaton reads a neutral symbol $c \in \Sigma_0$ at the outer level of brackets,
it simulates $\mathcal{M}$.
\begin{equation*}
	\delta'_c(p) = \eta_c(p)
\end{equation*}
If a left bracket ${<} \in \Sigma_{+1}$ is read at the outer level of brackets,
the new automaton begins simulating all $m$ input-driven automata
on the substring beginning with this bracket,
whereas the bracket is pushed onto the stack
along with the current state $p \in P$.
\begin{align*}
	\delta'_<(p) &= \big(\delta_<(q^{(1)}_0), \ldots, \delta_<(q^{(m)}_0)\big)
		\\
	\gamma'_<(p) &= (p, {<})
\end{align*}
Inside the brackets, the $m$ input-driven automata are simulated independently of each other.
\begin{align*}
	\delta'_c\big(q^{(1)}, \ldots, q^{(m)}\big) &= \big(\delta_c(q^{(1)}), \ldots, \delta_c(q^{(m)})\big)
		\\
	\delta'_<\big(q^{(1)}, \ldots, q^{(m)}\big) &= \big(\delta_<(q^{(1)}), \ldots, \delta_<(q^{(m)})\big)
		\\
	\gamma'_<\big(q^{(1)}, \ldots, q^{(m)}\big) &= \big(\gamma_<(q^{(1)}), \ldots, \gamma_<(q^{(m)})\big)
		\\
	\delta'_>\big((q^{(1)}, \ldots, q^{(m)}), (s^{(1)}, \ldots, s^{(m)})\big) &=
		\big(\delta_>(q^{(1)}, s^{(1)}), \ldots, \delta_>(q^{(m)}, s^{(m)})\big)
\end{align*}
Eventually, a right bracket at the outer level is read,
and the new automaton pops a pair $(p, {<})$.
Let the current state of the automaton be $(q^{(1)}, \ldots, q^{(m)})$.
The matching left bracket ($<$) was read by the simulated automata
in the states $\delta_<(q^{(1)}_0), \ldots, \delta_<(q^{(m)}_0)$.
Then, after reading the right bracket ${>} \in \Sigma_{-1}$,
each $i$-th simulated automaton is in the state
$\widehat{q}^{(i)}=\delta_>(q^{(i)}, \gamma(q^{(i)}_0, {<}))$.
Since the languages $K_1, \ldots, K_m$ are pairwise disjoint,
at most one of these automata can be in an accepting state.
If the $i$-th automaton accepts,
this indicates that a well-nested string $x \in K_i$ has just been read,
and the simulated Muller automaton
accordingly makes the transition by the symbol $a_i$;
if none of the input-driven automata accept,
the new automaton's entire computation fails.
\begin{equation*}
	\delta'_>(p, {<}) = \begin{cases}
		\eta(p, a_i),
			& \text{if } \delta_>(q^{(i)}, \gamma(q^{(i)}_0, {<})) \in F_i
		\\
		p_{dead},
			& \text{if there is no such $i$}
	\end{cases}
\end{equation*}
The Muller acceptance conditions for the new automaton
are defined to check that the set of states in $P$
visited on the outer level of brackets infinitely often
is in the set $\mathcal{F}$,
whereas the states visited inside the brackets do not affect acceptance.
\begin{equation*}
	\mathcal{F}'=\set{F \subseteq Q'}{F \cap P \in \mathcal{F}}
\end{equation*}
\end{proof}

A variant of the same construction produces an automaton
with a B\"uchi acceptance condition.

\begin{lemma}\label{buchi_regular_characterization_to_buchi_nidpda_lemma}
Let $K_1, \ldots, K_m \subseteq \ewn$ be pairwise disjoint languages,
each recognized by an input-driven automaton,
and let $M$ be a regular $\omega$-language over the alphabet $\Omega=\{a_1, \ldots, a_m\}$.
Then, the following language is recognized by a nondeterministic input-driven B\"uchi automaton.
\begin{equation*}
	L = \set{x_1 x_2 \ldots x_j \ldots}{x_j \in K_{i_j} \text{ for each } j \geqslant 0, \:
		a_{i_1} a_{i_2} \ldots a_{i_j} \ldots \in M}
\end{equation*}
\end{lemma}
\begin{proof}[Sketch of a proof.]
This time, the proof begins with taking 
a \emph{nondeterministic B\"uchi} automaton
$\mathcal{B}=(\Omega, P, p_0, \eta, F)$
recognizing $M$.
As in Lemma~\ref{muller_regular_characterization_to_muller_didpda_lemma},
let $\mathcal{A}_i=(\Sigma, Q_i, \Gamma_i, q^{(i)}_0, \langle\delta^{(i)}\rangle_{c \in \Sigma}, F_i)$,
be an IDPDA recognizing $K_i$.

A nondeterministic input-driven B\"uchi automaton $\mathcal{B}'$
uses the same states as $\mathcal{M}'$
in Lemma~\ref{muller_regular_characterization_to_muller_didpda_lemma},
and has basically the same transitions as $\mathcal{M}'$,
except for using nondeterminism on the outer level of brackets
whenever so does the automaton $\mathcal{B}$ being simulated.
Its set of accepting states under the B\"uchi acceptance condition
is the same as in $\mathcal{B}$.
\end{proof}

Putting the above results together
completes the characterization of well-nested $\omega$-input-driven languages
by $\omega$-regular languages,
and also establishes the equivalence
of several kinds of input-driven automata
on well-nested infinite strings.

\begin{theorem}\label{main_theorem}
Let $\Sigma=\Sigma_{+1} \cup \Sigma_{-1} \cup \Sigma_0$ be an alphabet.
Let $L \subseteq \Sigma^\omega$ be a language of infinite strings
and assume automata operating only on well-nested infinite strings.
Then the following three conditions are equivalent:
\begin{enumerate}
\item\label{main_theorem_detM}
	$L$ is recognized by a deterministic input-driven Muller automaton;
\item\label{main_theorem_nondetM}
	$L$ is recognized by a nondeterministic input-driven Muller automaton;
\item\label{main_theorem_nondetB}
	$L$ is recognized by a nondeterministic input-driven B\"uchi automaton;
\item\label{main_theorem_condition}
	there is a finite set of pairwise disjoint languages,
	$K_1, \ldots, K_m \subseteq \Sigma_0 \cup \Sigma_{+1} \mathrm{wn}(\Sigma) \Sigma_{-1}$,
	each recognized by an input-driven automaton,
	and a regular $\omega$-language $M$ over the alphabet $\Omega=\{a_1, \ldots, a_m\}$
	such that $L$ has the following representation.
	\begin{equation*}
		L = \set{x_1 x_2 \ldots x_j \ldots}{a_{i_1} a_{i_2} \ldots a_{i_j} \ldots \in M,
			\; x_j \in K_{i_j} \text{ for each } j \geqslant 0}
	\end{equation*}
\end{enumerate}
\end{theorem}
\begin{proof}
\textbf{($\ref{main_theorem_detM} \Rightarrow \ref{main_theorem_nondetM}$)}
Trivial.

\textbf{($\ref{main_theorem_nondetB} \Rightarrow \ref{main_theorem_nondetM}$)}
Trivial.

\textbf{($\ref{main_theorem_nondetM} \Rightarrow \ref{main_theorem_condition}$)}
By Lemma~\ref{muller_nidpda_by_muller_regular_characterization_lemma}.

\textbf{($\ref{main_theorem_condition} \Rightarrow \ref{main_theorem_detM}$)}
By Lemma~\ref{muller_regular_characterization_to_muller_didpda_lemma}.

\textbf{($\ref{main_theorem_condition} \Rightarrow \ref{main_theorem_nondetB}$)}
By Lemma~\ref{buchi_regular_characterization_to_buchi_nidpda_lemma}.
\end{proof}

The exact complexity of the transformation between models
is left as a question for future research.

\section{On the Wadge-Wagner hierarchy}\label{section_wadge}

Here we apply the results of Section~\ref{section_characterization}
to relate the Wadge degrees of IDPDA-recognisable languages in $\wnomega$ (w.r.t. the topology induced by the Cantor topology on $\Sigma^\omega$)
to those of regular $\omega$-languages.

Recall that for subsets $L,M\subseteq X$ of a topological space $X$,
the subset $L$ is {\em Wadge reducible} to $M$ (denoted by $L\leq^X_WM$)
if $L=f^{-1}(M)$ for some continuous function $f$ on $X$.
The sets $L,M$ are {\em Wadge equivalent} ($L\equiv^X_WM$)
if $L\leq^X_WM$ and $M\leq^X_WL$.
The equivalence classes under the Wadge equivalence
are known as {\em Wadge degrees} in $X$. More generally, one can define Wadge reducibility
between subsets $L\subseteq X$ and $M\subseteq Y$
of different topological spaces $X,Y$ as follows:
$L\leq_WM$, if $L=f^{-1}(M)$ for some continuous function $f \colon X \to Y$. Recall that the {\em Borel sets} in $X$ are generated from the open sets by repeated applications of complement and countable intersection. The Borel sets are organised in the Borel hierarchy; in particular, the $\mathbf{\Pi}_2$-sets in $X$ are the countable intersections of Boolean combinations of open sets.

The structure of Wadge degrees of Borel sets was first introduced and studied in \cite{wad} for the Baire space $X=\omega^\omega$ of infinite words over a countable alphabet. A remarkable fact is that this structure is very simple (almost well ordered) which leads to an elegant topological classification of the Borel sets. With minor modifications, the results for the Baire space also hold for the Cantor space $X=\Sigma^\omega$ (the open sets for this space are sets of the form $U\cdot\Sigma^\omega$, $U\subseteq\Sigma^*$). This is of interest for theoretical computer science because
one can consider Wadge degrees of $\omega$-languages over $\Sigma$
recognized by different kinds of automata and obtain topological classifications of the corresponding $\omega$-languages.

This direction was initiated by Wagner~\cite{wag} who characterised the structure $(\mathcal{R}_\Sigma;\leq_W)$ of Wadge degrees of regular $\omega$-languages over $\Sigma$
now known as the \emph{Wagner hierarchy}.
This result also holds for the structure $(\mathcal{R};\leq_W)$
of Wadge degrees of regular $\omega$-languages
over arbitrary finite alphabets: for any finite non-unary alphabet $\Omega$ we have $(\mathcal{R};\leq_W)\simeq(\mathcal{R}_\Omega;\leq_W)$,
where the relation $\simeq$ between preorders
means isomorphism of the corresponding quotient-posets.

Later, Wadge degrees of several important classes of $\omega$-languages were investigated. It was shown, in particular, that Wadge degrees of push down $\omega$-languages are much richer than those of regular $\omega$-languages (see e.g. \cite{dup,fin} and references therein).
In the context of our paper, a natural question is to characterise the structure $(\omega\textrm{-IDPDA}(\Sigma);\leq_W)$ of Wadge degrees of languages recognized by input-driven automata.
A theorem by L\"oding et al.~\cite[Thm.~6]{LoedingMadhusudanSerre}
implies that this structure is larger than $(\mathcal{R};\leq_W)$,
and that $(\mathcal{R};\leq_W)\not\simeq(\omega\textrm{-IDPDA}(\Sigma);\leq_W)$. We postpone the discussion of  the structure $(\omega\textrm{-IDPDA}(\Sigma);\leq_W)$  till a next publication because it requires  techniques different from those we used in this paper. 

Here, we formulate some immediate corollaries of the results of Section~\ref{section_characterization} to the relevant structure $(\wnID;\leq_W^X)$, where $\wnID$ is the class of languages of well-nested $\omega$-words recognised by the IDPDA, and $\leq_W^X$ is the Wadge reducibility in the space $X=\wnomega$ (considered as a subspace of $\Sigma^\omega$).
Namely, we show that $(\mathcal{R};\leq_W)\simeq(\wnID;\leq_W^X)$,
for any non-unary alphabet $\Sigma$. Note that the structure $(\wnID;\leq_W^X)$ is different from $(\wnID;\leq_W)$ where $\leq_W$ is the Wadge reeducibility on $\Sigma^\omega$. The Wadge reducibility in ``natural'' spaces was studied by several authors (see e.g. \cite{mss15,du20} and references therein). We start with some lemmas.

\begin{lemma}\label{lem0} 
The set $\wnomega$ is a $\mathbf{\Pi}_2$-subset of $\Sigma^\omega$ homeomorphic to the Baire space.
 \end{lemma}

\begin{proof}
Considering $\ewn$ as a countable alphabet, we may think that $\ewn^\omega$ is the Baire space. By the remarks in the beginning of Section \ref{idomega}, the infinite concatenation $h:\ewn^\omega\to\Sigma^\omega$ is a bijection between $\ewn^\omega$ and $\wnomega$. Comparing the topologies on $\ewn^\omega$ and on $\Sigma^\omega$ we see that $h$ is a homeomorphism between $\ewn^\omega$ and the subspace $\wnomega$ of $\Sigma^\omega$. 

It remains to show that $\wnomega$ is a $\mathbf{\Pi}_2$-subset of $\Sigma^\omega$. This immediately follows from the obvious remark that, for each $\alpha\in\Sigma^\omega$, the condition $\alpha\in\wnomega$ is equivalent to $\forall i\exists u\in \wn(\alpha[i]\sqsubseteq u\sqsubseteq\alpha)$, where $\alpha[i]$ is the prefix of $\alpha$ of length $i$ and $\sqsubseteq$ is the prefix relation.
 \end{proof}
 
\begin{lemma}\label{lem1} 
For any $L\in\wnID$
there exists a regular $\omega$-language $M$ over a finite alphabet $\Omega=\{a_1, \ldots, a_m\}$
such that $L\equiv_WM$.
\end{lemma}
\begin{proof}
Let $K_1, \ldots, K_m$, $\Omega=\{a_1, \ldots, a_m\}$, and $M$
be as in Lemma~\ref{muller_idpda_by_muller_regular_characterization_lemma}.
Without loss of generality, one can assume that every $K_i$ is non-empty
(otherwise, it is omitted from the list)
and that $K_1\cup\cdots\cup K_m=\ewn$
(otherwise, the language $K_{m+1}=\ewn\setminus(K_1\cup\cdots\cup K_m)$ is added to the list,
and the letter $a_{m+1}$ is added to $\Omega$).

Define $f \colon \wnomega\to \Omega^\omega$ by $f(x_1x_2\cdots)=a_{i_1} a_{i_2} \cdots$,
where, for each $j\geq1$, $i_j$ is the unique element of $\{1,\ldots,m\}$ with $x_j\in K_{i_j}$.
Then $f$ is continuous and $L=f^{-1}(M)$, so $L\leq_WM$ via $f$.
Define $g \colon \Omega^\omega\to\wnomega$ by $g(a_{i_1} a_{i_2} \cdots)=x_{i_1} x_{i_2} \cdots$,
where, for each $j\in\{1,\ldots,m\}$, $x_j$ is a fixed element of $K_j$.
Then $g$ is continuous and $M=f^{-1}(L)$, so $M\leq_WL$ via $g$. Therefore, $M\equiv_WL$.
 \end{proof}
 
\begin{lemma}\label{lem2} 
Let $m\geq1$, $\Omega=\{a_1, \ldots, a_m\}$,
and let $\{K_1,\ldots,K_m\}$ be a partition of $\ewn$
into non-empty languages recognized by input-driven automata.
Then, for every regular $\omega$-language $M$ over $\Omega$,
there exists $L \in \wnID$ such that $L\equiv_WM$.
\end{lemma}
\begin{proof}
Let $f \colon \wnomega\to \Omega^\omega$ be as in the proof of Lemma~\ref{lem1}.
By Lemma~\ref{muller_regular_characterization_to_muller_didpda_lemma}, for the language
$L=f^{-1}(M)$ we have $L\equiv_WM$. 
\end{proof}

\begin{theorem}\label{theor} 
For any non-unary  $\Sigma=\Sigma_{+1} \cup \Sigma_{-1} \cup \Sigma_0$ we have:
$(\mathcal{R};\leq_W)\simeq(\wnID\leq^X_W)$.
 \end{theorem}
\begin{proof}
Associate with any $L\in\wnID$
a language $M\in\mathcal{R}$ as in Lemma~\ref{lem1},
so in particular $L\equiv_WM$.
Then $L\leq L'$ clearly implies $M\leq_WM'$ for all $L,L'\in\wnID$.

Now we show that for any $M\in\mathcal{R}$ there exists $L\in\wnID$
such that $L\equiv_WM$.
Let $\Omega=\{a_1, \ldots, a_m\}$ be a finite alphabet with $M\in\mathcal{R}_\Omega$.
Since $\Sigma$ is non-unary,
there is clearly a partition $\{K_1,\ldots,K_m\}$ of $\ewn$ into non-empty languages
recognized by input-driven automata. By the proof of Lemma~\ref{lem2}, we can take $L=f^{-1}(M)$.

Returning to the map $L\mapsto M$ from the beginning of the proof,
it remains to show that $M\leq_WM'$ implies $L\leq_WL'$.
Let $\Omega$ be any non-unary finite alphabet.
By remarks in the beginning of this section,
there are $M_1,M'_1\in\mathcal{R}_\Omega$ such that $M\equiv_WM_1$ and $M'\equiv_WM'_1$.
By the preceding paragraph, there exist $L_1,L'_1\in\wnID$
such that $L_1\equiv_WM_1$ and $L'_1\equiv_WM'_1$.
Since also $L\equiv_WL_1$ and $L'\equiv_WL'_1$, it follows that $L\leq_WL'$. Therefore, $L\mapsto M$ induces a desired isomorphism between the quotient-posets of $(\wnID\leq^X_W)$ and $(\mathcal{R};\leq_W)$.
 \end{proof}


\begin{thebibliography}{99}

\bibitem{AlurMadhusudan} R. Alur, P. Madhusudan,
	\href{http://dx.doi.org/10.1145/1007352.1007390}
	{``Visibly pushdown languages''},
	\emph{ACM Symposium on Theory of Computing}
	(STOC 2004, Chicago, USA, 13--16 June 2004),
	202--211.

\bibitem{AlurMadhusudan_JACM} R. Alur, P. Madhusudan,
	\href{http://dx.doi.org/10.1145/1516512.1516518}
	{``Adding nesting structure to words''},
	\emph{Journal of the ACM},
	56:3 (2009).

\bibitem{vonBraunmuehl_Verbeek} B. von Braunm\"uhl, R. Verbeek,
	\href{http://dx.doi.org/10.1016/S0304-0208(08)73072-X}
	{``Input driven languages are recognized in $\log n$ space''},
	\emph{Annals of Discrete Mathematics},
	24 (1985), 1--20.

\bibitem{dup} J. Duparc,
	{``A hierarchy of deterministic context-free $\omega$-languages''},
	\emph{Theoretical Computer Science},
	290:3 (2003), 1253--1300.

\bibitem{du20} J. Duparc, L. Vuilleumier,
	{``The Wadge order on the Scott domain is not a well-quasi-order''},
	\emph{Journal of Symbolic Logic}, 85:1 (2020), 300--324.

\bibitem{fin} O. Finkel,
	{``Topological complexity of context-free $\omega$-languages: a survey''},
	In: Language, Culture, Computation: Studies in Honor of Yaacov Choueka.
Ed. by Nachum Dershowitz and Ephraim Nissan, Part I, Computing, Theory and Technology,
	LNCS 8001, Springer, 2014.

\bibitem{LoedingMadhusudanSerre} C. L\"oding, P. Madhusudan, O. Serre,
	\href{https://doi.org/10.1007/978-3-540-30538-5_34}
	{``Visibly pushdown games''},
	FSTTCS 2004, 408--420.

\bibitem{Mehlhorn} K. Mehlhorn,
	\href{http://dx.doi.org/10.1007/3-540-10003-2_89}
	{``Pebbling mountain ranges and its application to DCFL-recognition''},
	\emph{Automata, Languages and Programming}
	(ICALP 1980, Noordweijkerhout, The Netherlands, 14--18 July 1980),
	LNCS 85, 422-435.

\bibitem{mss15} L. Motto Ros, P. Schlicht, V. Selivanov,
	{``Wadge-like reducibilities on arbitrary quasi-Polish spaces''},
	\emph{Mathematical Structures in Computer Science}, 25, Special Issue 08 (2015), 1705--1754.

\bibitem{idpda_sigact_survey} A. Okhotin, K. Salomaa,
	\href{http://doi.acm.org/10.1145/2636805.2636821}
	{``Complexity of input-driven pushdown automata''},
	\emph{SIGACT News},
	45:2 (2014), 47--67.

\bibitem{vpda_sc_unamb} A. Okhotin, K. Salomaa,
	\href{http://dx.doi.org/10.1016/j.tcs.2014.11.015}
	{``Descriptional complexity of unambiguous input-driven pushdown automata''},
	\emph{Theoretical Computer Science},
	566 (2015), 1--11.

\bibitem{vpda_sc} A. Okhotin, K. Salomaa,
	\href{http://dx.doi.org/10.1016/j.jcss.2017.02.001}
	{``State complexity of operations on input-driven pushdown automata''},
	\emph{Journal of Computer and System Sciences},
	86 (2017), 207--228.

\bibitem{wad} W. Wadge. {\em Reducibility and determinateness in the
Baire space.} PhD thesis, University of California, Berkely, 1984.

\bibitem{wag} K. Wagner, ``On $\omega$-regular sets'',
	{\em Information and Control}, 43 (1979), 123--177.

\end{thebibliography}
\end{document}